\documentstyle[preprint,aps]{revtex}
\begin{document}
\tightenlines
\draft

\title{QED radiative corrections to parity nonconservation in heavy atoms}

\author{M.Yu.Kuchiev
\thanks{kuchiev@newt.phys.unsw.edu.au} and V.V. Flambaum
\thanks{flambaum@newt.phys.unsw.edu.au} }
\address{School of Physics,
University of New South Wales,Sydney 2052,  Australia}
\maketitle

\begin{abstract}
The self-energy and vertex QED radiative corrections ($\sim Z \alpha^2 f(Z
\alpha)$) are shown to give a large negative contribution to the parity
nonconserving (PNC) amplitude in heavy atoms.  The correction $-0.7(2)\%$
found for the $6s-7s$ PNC amplitude in $^{133}$Cs brings the experimental
result for this transition into agreement with the standard model.  The
calculations are based on a new identity that expresses the radiative
corrections to the PNC matrix element via corrections to the energy shifts
induced by the finite nuclear size.  \end{abstract}


It has been discovered recently that there exists a consistent deviation of
experimental data on parity nonconservation (PNC) in atoms from predictions
of the standard model. This paper demonstrates that this contradiction is
removed by the self-energy and vertex QED radiative corrections, which
prove to be much larger than anticipated. The corrections are evaluated with
the help of a new identity that expresses them via similar radiative
corrections to the energy shifts induced by the finite nuclear size.

Experimental investigation  of the $6s-7s$ PNC amplitude in $^{133}$Cs
initiated by Bouchiat and Bouchiat \cite{bouchiat}, was carried further by
Gilbert and Wieman \cite{gilbert_wieman_86}, and by Wood {\it et al}
\cite{wood_97} who reduced the error to $0.3\%$, sparking an interest in
the atomic PNC calculations that are crucial for the analysis of the
experimental data. Accurate previous calculations of Refs.
\cite{dzuba_89,blundell_92} have recently been revisited by Kozlov {\it et
al} \cite{kozlov_01} and Dzuba {\it et al} \cite{dzuba_01,dzuba_02}.
Bennett and Wieman \cite{bennett_wieman_99} analyzed the theoretical data
\cite{dzuba_89,blundell_92}, comparing it with available experimental data
on dipole amplitudes, polarizabilities and hyperfine constants for Cs, and
suggested that the theoretical error for the PNC amplitude should be
reduced from $1\%$ to $0.4 \%$. It has been recognized recently that
several, previously neglected  phenomena contribute at the required level
of accuracy. Derevianko \cite{derevianko_00} found that the Breit
corrections give $-0.6 \%$, the value confirmed in
\cite{dzuba_harabati_01,kozlov_01}. Sushkov \cite{sushkov_01} pointed out
that the radiative corrections may be comparable with the Breit
corrections. The calculations of Johnson {\it et al} \cite{johnson_01}
demonstrated that indeed, the QED vacuum polarization gives $0.4 \%$, the
value confirmed in \cite{milstein_sushkov_01,dzuba_01,kf_02}.

Ref.  \cite{bennett_wieman_99} indicates that there is a $2.3\sigma$
deviation of the weak charge $Q_{\mathrm W}$ extracted from the atomic PNC
amplitude \cite{wood_97} from predictions of the standard model
\cite{groom_00}. More recent works \cite{johnson_01,dzuba_02}, in which
the Breit corrections ($-0.6 \%$) and the  QED vacuum polarization ($0.4
\%$) were included, give similar deviations  $2.2\sigma$ and $2.0\sigma$
respectively. We show that this contradiction is removed by the self-energy
and vertex radiative corrections. The corrections of this type were
considered previously by Marciano and Sirlin \cite{marciano_sirlin_83} and
Lynn and Sandars \cite{lynn_sandars_94} using the plane wave  approximation
that resulted in a small value $\sim 0.1 \%$.  The expectation of
Ref.\cite{johnson_01} is that the Coulomb field of the atomic nucleus
should not produce any drastic effect on these radiative corrections, which
should remain small. This assessment is supported  by Ref.
\cite{milstein_sushkov_01} that mentions in passing preliminary results of
calculations indicating that the self-energy and vertex  corrections
altogether are small. The opposite conclusion of Ref. \cite{dzuba_01},
namely that the self-energy correction may give a substantial contribution,
was not decisive, because the model approach pursued in this work was not
gauge invariant.

Let us show that there is a precise relation that expresses the QED
radiative corrections to the PNC matrix element via similar radiative
corrections to the energy shifts of the atomic electron induced by the
finite nuclear size (FNS). This relation can be presented as

\begin{equation}\label{ddd}
\delta_{\mathrm PNC,\,sp} = \frac{1}{2}\,  ( \delta_{ \mathrm FNS,\,s } +
\delta_{\mathrm FNS,\,p })~,
\end{equation}
where $\delta_{ \mathrm PNC,\, sp }$ is the relative radiative  correction
to the PNC matrix element between $s_{1/2}$ and $p_{1/2}$ orbitals. The
term {\em relative correction} used above indicates that the
correction is divided by the matrix element itself

\begin{equation}\label{d}
\delta_{\mathrm PNC,\,sp} = \frac{\langle \psi_{s,1/2} | H_{\mathrm PNC}
|\psi_{p,1/2}\rangle^ {{\mathrm rad} } }  {   \langle \psi_{s,1/2} |
H_{\mathrm PNC} |\psi_{p,1/2}\rangle~~~ }~,
\end{equation}
Similarly $\delta_{ {\mathrm FNS},\,s}$ and $\delta_{{\mathrm FNS},\,p}$
are the relative radiative corrections to the FNS energy shifts
$E_{\mathrm FNS,\,s}, ~E_{\mathrm FNS,\,p},$ for the the chosen $s_{1/2}$
and $p_{1/2}$ electron states

\begin{eqnarray}\label{dd}
\delta_{ {\mathrm
FNS},\,l} =  E_{ {\mathrm FNS},\,l}^{\mathrm rad}/ E_{{\mathrm FNS},\,l} ~,
\quad  l = s, p ~.
\end{eqnarray}
The operator $H_{\mathrm PNC}$ in (\ref{d}) describes the PNC part of the
electron Hamiltonian induced by the $Z$-boson exchange

\begin{equation}\label{05}  H_{\mathrm PNC} = (
2\sqrt{2} )^{-1}G_{\mathrm F} Q_{\mathrm W}\,  \rho(r)\,\gamma_5 ~.
\end{equation}
Here $G_{\mathrm F}$ and $ Q_{\mathrm W} $ are the Fermi constant and the
nuclear weak charge, and $\rho(r)$ is the nuclear density. The FNS energy
shifts can be presented as matrix elements of the potential  $\delta
V_{\mathrm FNS}(r)$, which describes the deviation of the nuclear potential
from the pure Coulomb one, $E_{ {\mathrm FNS},\,l } = \langle \psi_{
l,1/2 } | \delta V_{\mathrm FNS} |\psi_{l,1/2}\rangle~,l=s,p$. Equality
(\ref{ddd}) may be established for the sum of all QED radiative
corrections ($\sim Z \alpha^2 f(Z \alpha)$), or specified for any gauge
invariant class of them. We concentrate  our attention on the self-energy
and vertex corrections in the lowest  order of the perturbation theory
described by the Feynman diagrams in  Fig.  \ref{one}, calling them the
e-line corrections, though the vacuum polarization is also briefly
discussed below. The vertex in diagrams of Fig.  (\ref{one}) originates
from the PNC Hamiltonian (\ref{05}), the left and right external legs
describe the wave functions $ \psi_{s,1/2}({\bf r})$ and $\psi_{p,1/2}({\bf
r})$ for the considered atomic states.  The sum of these diagrams is gauge
invariant, though each one of them is not. We can use this fact for our
advantage, choosing a gauge in which the vertex correction Fig.
{\ref{one}(a) is zero. To see that this is possible we can consider, for
example, the gauge in which the photon propagator is $D^{\mu\nu}(k) =
(g^{\mu\nu} - \xi f(k) k^\mu k^\nu)/k^2$, where $f(k)$ is some nonzero,
nonsingular function of $k$, and  tune one parameter $\xi$ to annihilate
the vertex correction for the  chosen pair of atomic $s_{1/2},p_{1/2}$
levels. In this gauge only the self-energy corrections Fig. \ref{one}(b,c)
contribute to the PNC transition between this pair of states.

The PNC Hamiltonian (\ref{05}) is localized inside the nuclear interior
$r\le R$, where $R$ is the nuclear radius. In contrast, the radiative
corrections take place at separations comparable with the Compton radius $r
\sim m^{-1}$ which is much bigger than the nucleus, $ mR \ll 1 $
(relativistic units $\hbar=c=1$ are used, if not stated  otherwise).  This
difference of the two scales allows us to simplify the problem. Consider
the diagram Fig.  \ref{one}(b) in which the initial wave function (the
left leg) is $\psi_{s,1/2}({\bf r})$.  The radiative correction induced by
the self-energy operator results in a variation of this wave function
$\delta^{\mathrm rad} \psi_{s,1/2}({\bf r})$ which we need to evaluate at
the nucleus, where the weak interaction takes place. In this region the
shape of the function does not change, because the perturbation caused by
the radiative correction is localized far away. Thus, inside the nucleus
$\delta^{\mathrm rad} \psi_{s,1/2}({\bf r}) = C_s^{\mathrm rad}
\psi_{s,1/2}({\bf r})$, where $C_s^{\mathrm  rad}$ is an ${\bf
r}$-independent factor. Similarly, for the right leg  (diagram Fig.
\ref{one}(c))  $\delta^{\mathrm rad} \psi_{p,1/2}({\bf r}) = C_p^{\mathrm
rad} \psi_{p,1/2}({\bf r})$. Using these variations of the wave function
we express the relative radiative correction to the PNC matrix element
(\ref{d}) in terms of the factors $C_s^{\mathrm rad},~C_p^{\mathrm rad}$

\begin{equation}\label{FF}
\delta_{\mathrm PNC,\,sp} = C_s^{\mathrm rad} +C_p^{\mathrm rad}~.
\end{equation}
We took into account here that in the chosen gauge $\langle \psi_{s,1/2} |
H_{\mathrm PNC} |\psi_{p,1/2}\rangle^{\mathrm rad} =  \langle
\delta^{\mathrm rad} \psi_{s,1/2} | H_{\mathrm PNC} |\psi_{p,1/2}\rangle  +
\langle \psi_{s,1/2} | H_{\mathrm PNC} |\delta^{\mathrm rad}\psi_{p,1/2}
\rangle$.

Let us discuss now the e-line radiative corrections to the FNS energy shift
for the $s_{1/2}$ level that are described by the same Feynman diagrams in
Fig.\ref{one} in which the vertex is given by deviation of the potential
from the pure Coulomb one $\delta V_{\mathrm FNS}(r)$. We can again choose
the gauge in which the vertex correction is zero for the given $s_{1/2}$
state.  Moreover, we can assume that the vertex radiative corrections are
zero simultaneously for the FNS energy shift and for the PNC matrix element
(for the chosen pair of $s_{1/2},p_{1/2}$ states). This is possible
because gauge transformations include an infinite number of parameters.
The gauge $D^{\mu\nu}(k) = (g^{\mu\nu} - f(k) k^\mu k^\nu)/{k^2}$ presents
them via an arbitrary function $f(k)$ that can be chosen to satisfy the
conditions formulated above.
In this gauge the radiative correction to $E_{\mathrm FNS,\,s} =
\langle \psi_{s,1/2} | \delta V_{\mathrm FNS} |\psi_{s,1/2}\rangle$  is
expressed via the variation of the wave function $\delta^{\mathrm
rad}\psi_{s,1/2}({\bf r})$, which is essential only inside the nucleus
where $\delta V_{\mathrm FNS}(r)$ is located. Thus, using arguments similar
to the ones that led us to (\ref{FF}), we find the relative correction to
the FNS energy shift

\begin{equation}\label{2Fs}
\delta_{ {\mathrm  FNS},\,s} = 2 C_s^{\mathrm rad}~,
\end{equation}
where  the factor 2 accounts for two diagrams (b) and (c) in Fig.\ref{one}
that give identical contributions.  Similarly for the $p_{1/2}$ level,

\begin{equation}\label{2Fp}
\delta_{ {\mathrm FNS},\,p} = 2 C_p^{\mathrm rad}~.
\end{equation}
The factors $C_s^{\mathrm rad},~C_p^{\mathrm rad}$ are determined
by the radiative corrections, being independent on the nature of a
perturbative operator localized on the nucleus. They have, therefore, the
same values for the PNC matrix element (\ref{FF}) and FNS energy shifts
(\ref{2Fs}),(\ref{2Fp}).  Combining these equations we immediately derive
Eq.(\ref{ddd}). The only parameter which governs the accuracy of the
presented derivation is the smallness of the nucleus compared with
the Compton radius. This makes Eq.(\ref{ddd}) a very accurate identity.
Similarly we derive this identity for the contribution of the QED vacuum
polarization. The result can be compared with Ref. \cite{kf_02} that
presents explicit variations of $s_{1/2}$ and $p_{1/2}$ wave functions  at
the origin induced by the vacuum polarization (see Eq.(43) of  Ref.
\cite{kf_02}). Using these wave functions to calculate corrections to the
PNC matrix element and FNS energy shifts, we find perfect agreement with
Eq.(\ref{ddd}).

Note that we do not consider here radiative corrections of the order  $\sim
\alpha/\pi$ which appear in the plane wave approximation. These
contributions have been included into the radiative corrections  to the
weak charge $Q_W$  (and the renormalization of the charge and electron mass
in the case of FNS energy shifts).  Correspondingly, we subtract the
contribution of the plane waves from Eq.(\ref{ddd}), considering only the
part of the corrections that depends on the atomic potential $\sim
Z\alpha^2 f(Z\alpha)$. For heavy atoms this subtlety is insignificant
numerically because the considered $Z$-dependent part of the correction is
bigger than the omitted $Z$-independent one, as we will see below.

Eq. (\ref{ddd}) presents the e-line corrections to the PNC matrix element,
which are difficult to calculate, in terms of the corrections to the FNS
energy shifts that have been well-studied both numerically, by Johnson
and Soff \cite{johnson_soff_85}, Blundell \cite{blundell_92}, Cheng {\em
et al} \cite{cheng_93} and Lindgren {\em et al} \cite{lindgren_93}, and
analytically, by Pachucki \cite{pachucki_93} and Eides and Grotch
\cite{eides_97}. Ref. \cite{cheng_93} presents the e-line radiative
corrections to the FNS energy shifts for $1s_{1/2}$, $2s_{1/2}$ and
$2p_{1/2}$ levels in hydrogenlike ions with atomic charges $Z=60,70,80,90$.
Eq. (\ref{ddd}) contains relative corrections, therefore we need to
calculate the FNS energy shifts $E_{\mathrm FNS}$. We did this by solving
the Dirac equation with the conventional Fermi-type nuclear distribution
$\rho(r) = \rho_0 /\{ 1 + \exp [(r-a)/c] \} $. Parameters $a,c$ were taken
the same as in \cite{cheng_93}, namely $a = 0.523$ fm and $c$ chosen to
satisfy $R_{\mathrm rms} = 0.836 A^{1/3} + 0.570$ fm. Using the results
of \cite{cheng_93} and this calculation we obtained the relative radiative
corrections shown in Fig. \ref{two}. In order to include the 
interesting case $Z=55$ and to account for all values of $55 \le Z \le 90$
we used interpolation formulae presented in \cite{cheng_93}. The relative
corrections for the $1s$ and $2s$ levels are roughly the same size. This
indicates that the radiative processes responsible for the correction take
place at separations smaller than the K-shell  radius, $r < Z \alpha
m^{-1}$, which is consistent with the assumption $r\sim m^{-1}$ above. For
these separations we can assume that, firstly, the screening of the nuclear
Coulomb field in many-electron atoms does not produce any significant
effect, and, secondly, the relative corrections does not depend on the
atomic energy level because for small separations all atomic
$ns_{1/2}$-wave functions exhibit similar behaviour. These arguments remain 
valid for the $p_{1/2}$ states as well, permitting us to presume 
that the results shown in Fig.  \ref{two}  
for the $2s_{1/2}$-levels and $2p_{1/2}$-levels
of hydrogenlike ions remain valid for $s_{1/2}$ and $p_{1/2}$ states of the
valence electron in a many-electron atom. We obtain the e-line radiative
corrections for the PNC matrix element using Eq.(\ref{ddd}) that expresses
them via the found corrections to the FNS energy shifts. 
The found PNC corrections, presented in Fig.
\ref{two} by the dotted line, are negative and large (much larger than the
neglected $Z$-independent part of the corrections).

Let us discuss the implications for the $6s-7s$ PNC amplitude in
$^{133}$Cs. The standard model value for the nuclear weak charge for Cs
\cite{groom_00} is

\begin{equation}\label{QW}
Q_W(^{133}{\mathrm Cs}) = \,-73.09 \,\pm\,(0.03)~.
\end{equation}
Ref. \cite{dzuba_02} refined previous calculations of Ref. \cite{dzuba_89}
extracting from the experimental PNC amplitude of Ref. \cite{wood_97}
the weak charge

\begin{equation}\label{72.18}
Q_W(^{133}{\mathrm Cs}) = \,-72.18\pm(0.29)_{\mathrm expt}\pm(0.36)_{\mathrm
theor}~,
\end{equation}
with the theoretical error $0.5\%$. It is consistent with
$Q_W(^{133}{\mathrm Cs}) = \,-72.21 \,\pm\,(0.28)_{\mathrm expt}\,\pm\,
(0.34)_{\mathrm theor}$ that was adopted in \cite{johnson_01} by taking the
average of the results of Refs.\cite{dzuba_89,blundell_92,kozlov_01}, and
accepting the theoretical error $0.4\%$ proposed in
\cite{bennett_wieman_99}. The weak charge in Eq.(\ref{72.18}) deviates from
the standard model (\ref{QW}) by $2.0\sigma$.

The e-line radiative correction derived from results presented in 
Fig. \ref{two} is $-0.73\pm(0.20) \%$, the error
reflects the uncertainty of the radiative corrections to the FNS energy
shift for the $2p_{1/2}$ level in Cs $E_{ {\mathrm FNS},2p,1/2}^{\mathrm
rad} = -0.0001(1)$ eV \cite{cheng_02} (more accurate value can, 
probably, be obtained by extrapolation of much
more reliable higher-$Z$ results shown in Fig. \ref{two}). 
Eq.(\ref{72.18}) combined with the e-line correction gives

\begin{equation}\label{72.71}  Q_W(^{133}{\mathrm Cs}) =  \,-72.71
\pm(0.29)_{\mathrm expt}\pm(0.39)_{\mathrm theor}~,
\end{equation}
which brings the PNC experimental amplitude of \cite{wood_97} within the
limits of the standard model (\ref{QW}). For heavier atoms the e-line
corrections become larger, while the error diminishes. 
For the Tl atom, which presents another interesting for applications case, 
we find the e-line correction $-1.61 \%$.

Relations similar to (\ref{ddd}) can be derived for any operator which is
localized at distances smaller than the Compton radius. One can even try to
apply it to the case of the hyperfine interaction (HFI), which has been
thoroughly investigated previously, see e.g.
\cite{blundell_97,sunnergren_98} and references therein, though the HFI has
a long-range tail $\sim 1/r^3$ that presents an obstacle for our method.
However, if convergence of the HFI matrix elements is fast, the relation
$\delta_{ {\mathrm FNS},\,s} \approx \delta_{ {\mathrm HFI},\,s}'$ should
hold. Here $\delta_{ {\mathrm HFI},\,s}'$ is the radiative correction to
the HFI for $s$-levels, the primed notation indicates that the
$Z$-independent Schwinger term $\alpha/(2\pi)$ should be excluded
(for heavy atoms this subtlety is not important.) Fig.\ref{two} shows the
e-line contribution to $\delta_{ {\mathrm HFI},\,s}'$ that was extracted
from \cite{blundell_97} using interpolation for all considered values of $Z$.
It agrees semi-quantitatively with $\delta_{ {\mathrm
FNS},\,s}$, deviation is less than 33 \%.  Overall, we observe that
$\delta_{ {\mathrm FNS},\,1s}, \delta_{ {\mathrm FNS},\,2s}, \delta_{
{\mathrm FNS},\,2p}$, and  $\delta_{ {\mathrm HFI},\,s}'$ all exhibit
similar behaviour, they are negative and large regardless of the
perturbative operator considered and quantum numbers of the wave
functions involved, which is in line with the main arguments of this paper.

In conclusion, large QED self-energy and vertex corrections to the parity
nonconservation amplitude in heavy atoms reconcile the experimental results
of Wood {\em et al} \cite{wood_97} in Cs with the standard model.

This work was supported by the Australian Research Council. The authors 
are thankful to K.T.Cheng for his calculations of the self-energy corrections
to the FNS energy shifts in Cs Ref. \cite{cheng_02}.

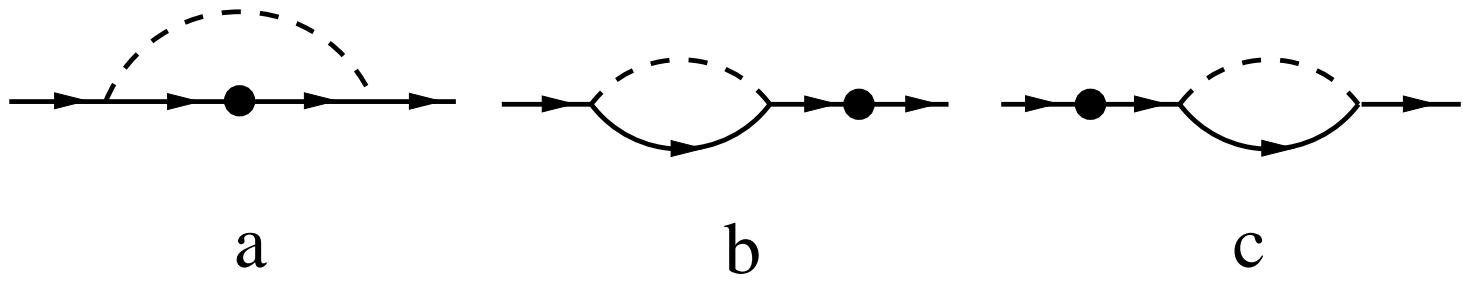
\begin{figure}
\caption{\label{one} The QED vertex (a) and self-energy (b),(c) corrections
to the PNC matrix element and the FNS energy shifts.  For the
electron-nucleus PNC interaction the vertex is given in Eq.~(\ref{05}), for
the FNS correction the vertex is produced by the short-range  potential
that describes the spreading of the nuclear charge.} \end{figure}

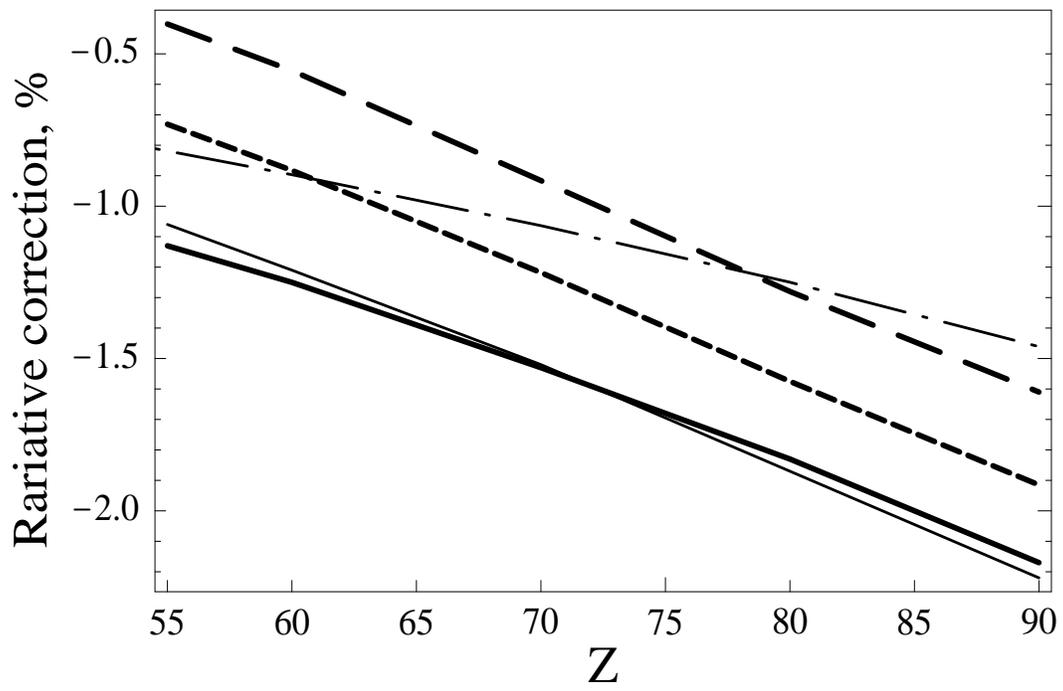
\begin{figure}
\caption{\label{two} The relative radiative corrections (in $\%$) induced
by the diagramms of Fig.1. Corrections to the FNS energy shifts for
$1s_{1/2},2s_{1/2}$, and $2p_{1/2}$ levels extracted from [21] as
discussed in the text are shown by thick, thin, and dashed lines. Dotted
line - predictions of Eq.(\ref{ddd}) for the radiative corrections to the
PNC matrix element. Dashed-dottet line - relative correction to the HFI
taken from [25].} \end{figure}


\begin{references}

\bibitem{bouchiat}
M.A.Bouchiat and C.Bouchiat J.Phys.  (Paris) {\bf 35}, 899 (1974);  {\bf
36}, 493 (1974).

\bibitem{gilbert_wieman_86}
S.L.Gilbert and C.E.Wieman, Phys.Rev.  A {\bf 34}, 792 (1986).

\bibitem{wood_97}
C.S.Wood, S.C.Bennett,  D.Cho, B.P.Masterson, J.L.Roberts,  C.E.Tanner, and
C.E.Wieman,  Science 275, 1759 (1997).

\bibitem{dzuba_89}
V.A.Dzuba, V.V.Flambaum, and O.P.Sushkov, Phys.  Lett A {\bf 141}, 147
(1989).

\bibitem{blundell_92}
S.A.Blundell, J.Sapirstein, and W.R.Johnson, Phys.  Rev. D {\bf 45}, 1602
(1992).

\bibitem{kozlov_01}
M.G.Kozlov, S.G.Porsev, and I.I.Tupitsyn, Phys. Rev.  Lett. {\bf 86}, 3260
(2001).

\bibitem{dzuba_01}
V.A.Dzuba, V.V.Flambaum, and J.S.M. Ginges, hep-ph/0111019 (2001).

\bibitem{dzuba_02}
V.A.Dzuba, V.V.Flambaum, and J.S.M. Ginges, hep-ph/0204134 (2002).

\bibitem{bennett_wieman_99}
S.C.Bennett and C.E.Wieman, Phys. Rev.  Lett.  {\bf 82}, 2484 (1999);  {\bf
82}, 4153 (1999);  {\bf 83},  889 (1999).

\bibitem{derevianko_00}
A.Derevianko, Phys. Rev. Lett. {\bf 85}, 1618 (2000).

\bibitem{dzuba_harabati_01}
V.A.Dzuba, C.Harabati, W.R.Johnson, and M.S.Safronova, Phys. Rev A {\bf
63}, 044103 (2001).

\bibitem{sushkov_01}
O.P.Sushkov, Phys. Rev. A, {\bf 63}, 042504 (2001).

\bibitem{johnson_01}
W.R.Johnson, I.Bednyakov, and G.Soff, Phys.  Rev.  Lett {\bf 87},  233001-1
(2001).

\bibitem{milstein_sushkov_01}
A.I.Milstein and O.P.Sushkov, hep-ph/0109257 (2001).

\bibitem{kf_02}
M.Yu.Kuchiev and V.V.Flambaum, hep-ph/0205012 (2002).

\bibitem{groom_00}
D.E.Groom {\it et al}, Eur.  Phys. J. C {\bf 15}, 1 (2000).

\bibitem{marciano_sirlin_83}
W.J.Marciano and A.Sirlin, Phys. Rev. D {\bf
27}, 552 (1983).

\bibitem{lynn_sandars_94}
B.W.Lynn and P.G.H.Sandars, J. Phys. B {\bf 27}, 1469 (1994).
\bibitem{johnson_soff_85}

W.R.Johnson and G.Soff, At. Data Nuc. Data Tables {\bf 33},  405 (1985).
\bibitem{blundell_92} S.A.Blundell, Phys. Rev. A {\bf 46}, 3762 (1992).

\bibitem{cheng_93}
K.T.Cheng, W.R.Johnson and J.Sapirstein, Phys. Rev A {\bf 47}, 1817 (1993).

\bibitem{lindgren_93}
I.Lindgren, H.Persson, S.Salomonson, and A.Ynnerman, Phys. Rev. A {\bf 47},
4555 (1993).


\bibitem{pachucki_93}
K.Pachucki, Phys.  Rev. A {\bf 48}, 120 (1993).

\bibitem{eides_97}
M.I.Eides and H.Grotch, Phys.  Rev. A {\bf 56}, R2507
(1997).

\bibitem{blundell_97}
S.A.Blundell, K.T.Cheng, and J.Sapirstein, Phys.  Rev. A {\bf 55}, 1857
(1997).

\bibitem{sunnergren_98}
P.Sunnergren, H.Persson, S.Salomonson, S.M.Sneider, I.Lindgren, and
G.Soff, Phys. Rev.  A {\bf 58}, 1055 (1998).

\bibitem{cheng_02}
K.T.Cheng. Private communication (2002).




\end{references}
\end{document}